\documentclass[aps,prd,twocolumn,epsf,nofootinbib,superscriptaddress]{revtex4}

\usepackage{graphicx,here}
\usepackage{amsmath,amssymb,latexsym}
\usepackage{bm}

\def\lsim{\mathrel{\raise.3ex\hbox{$<$\kern-.75em\lower1ex\hbox{$\sim$}}}}
\def\gsim{\mathrel{\raise.3ex\hbox{$>$\kern-.75em\lower1ex\hbox{$\sim$}}}}

\def\lbldef#1#2{\expandafter\gdef\csname #1\endcsname {#2}}

\def\href#1#2{#2}

\newcommand{\bwide}{\begin{widetext}}
\newcommand{\ewide}{\end{widetext}}
\newcommand{\beq}[1]{\begin{equation} \label{(#1)}}
\newcommand{\eeq}{\end{equation}}
\newcommand{\ba}[1]{\begin{eqnarray} \label{(#1)}}
\newcommand{\ea}{\end{eqnarray}}

\begin{document}
\hspace*{130mm}{\large \tt FERMILAB-PUB-10-021-A}

\title{Inelastic Dark Matter As An Efficient Fuel For Compact Stars}

\author{Dan Hooper}
\affiliation{Center for Particle Astrophysics, Fermi National Accelerator Laboratory, Batavia, IL 60510}
\affiliation{Department of Astronomy and Astrophysics, The University of Chicago, Chicago, IL  60637} 
\author{Douglas Spolyar}
\affiliation{Center for Particle Astrophysics, Fermi National Accelerator Laboratory, Batavia, IL 60510}
\author{Alberto Vallinotto}
\affiliation{Center for Particle Astrophysics, Fermi National Accelerator Laboratory, Batavia, IL 60510}
\author{Nickolay Y. Gnedin}
\affiliation{Center for Particle Astrophysics, Fermi National Accelerator Laboratory, Batavia, IL 60510}
\affiliation{Department of Astronomy and Astrophysics, The University of Chicago, Chicago, IL  60637} 
\affiliation{Kavli Institute for Cosmological Physics, The University of Chicago, Chicago, IL 60637}

\begin{abstract}

Dark matter in the form of weakly interacting massive particles is predicted to become gravitationally captured and accumulate in stars. While the subsequent annihilations of such particles lead to the injection of energy into stellar cores, elastically scattering dark matter particles do not generally yield enough energy to observably impact stellar phenomenology. Dark matter particles which scatter inelastically with nuclei (such that they reconcile the annual modulation reported by DAMA with the null results of CDMS and other experiments), however, can be captured by and annihilate in compact stars at a much higher rate. As a result, old white dwarf stars residing in high dark matter density environments can be prevented from cooling below several thousand degrees Kelvin. Observations of old, cool white dwarfs in dwarf spheroidal galaxies, or in the inner kiloparsec of the Milky Way, can thus potentially provide a valuable test of the inelastic dark matter hypothesis.

\end{abstract}


\maketitle

\section{Introduction}

Over the past two years, the DAMA/LIBRA collaboration has repeated their claim to have observed an annual modulation in their event rate, and have interpreted this as evidence of dark matter particles interacting with their detector at a confidence level of 8.2$\sigma$~\cite{dama}. Such an annual modulation in the rate of dark matter scattering events is predicted to arise as a result of the variation in the velocity of the Earth with respect to the Milky Way's dark matter halo. Both the reported amplitude and phase of DAMA's signal are consistent with a dark matter interpretation.

In typical dark matter models capable of producing the signal reported by DAMA, however, one would have expected corresponding signals to appear in other direct detection experiments. To the contrary, several such experiments, including CDMS~\cite{cdms}, CRESST~\cite{cresst}, CoGeNT~\cite{cogent}, XENON~\cite{xenon}, COUPP~\cite{coupp}, ZEPLIN~\cite{zeplin}, Edelweiss~\cite{edelweiss}, and KIMS~\cite{kims}, have reported null results, strongly constraining the dark matter interpretation of DAMA's annual modulation~\cite{compare}. This tension between DAMA and other direct detection experiments has motivated a number of scenarios in which dark matter particles scatter preferentially with the sodium iodide (NaI) detectors of DAMA/LIBRA relative to the materials used in other experiments (such as germanium and silicon used in CDMS). Such scenarios feature dark matter particles which interact with nuclei through a resonance~\cite{resonance}, interact with nuclei with a momentum dependence causing them to scatter more efficiently with NaI than other targets~\cite{ffdm}, or which interact with nuclei largely through inelastic processes~\cite{inelastic}. In particular, dark matter particles which, instead of elastically scattering with nuclei, scatter only through processes in which they are excited to a slightly ($\sim$$100$ keV) heavier state have generated a great deal of interest, and have been shown to be capable of reconciling the claims of the DAMA collaboration with the null results of other direct detection experiments~\cite{inelastic}.

In this article, we consider inelastic dark matter and study the role that such particle would play in compact stars, such as white dwarfs. In the case of dark matter candidates which scatter elastically with nuclei, their capture by and annihilation in stars is expected to provide an observable quantity of energy only in the most extreme environments, such as in a density spike of dark matter in the proximity of the supermassive black hole at the center of the Milky Way~\cite{burners}, or in the early phases of evolution of very massive population III stars~\cite{early}. Inelastic dark matter, in contrast, can potentially scatter with nuclei far more efficiently, but only if in possession of sufficient momentum to overcome the mass splitting between the ground and excited states. Because of the depth of the gravitational wells surrounding white dwarfs, inelastic dark matter particles will be accelerated to high velocities as they fall toward such a star, allowing them to easily accumulate enough momentum to scatter inelastically with the carbon or oxygen nuclei which make up the majority of the star's mass. As a result, inelastic dark matter particles will scatter in and be captured by white dwarf stars at very high rates. In their subsequent annihilations, inelastic dark matter particles can deposit a significant amount of energy into these stars, potentially increasing their luminosity and/or temperature. In particular, old and cool white dwarf stars that reside in regions of high dark matter density (such as in dwarf spheroidal galaxies, or in the inner regions of the Milky Way) may provide a valuable test of the inelastic dark matter scenario.

\section{Capture of Inelastic Dark Matter In White Dwarf Stars}

More than two decades ago, the rate at which elastically scattering dark matter particles are captured in the Sun was calculated~\cite{capture}. In many respects, the capture rate of inelastically scattering dark matter particles is similar. The key difference, however, is that in the inelastic scenario the dark matter-nucleus scattering cross section depends on the velocity of the incoming particle. Above the threshold for inelastic scattering, the cross section is given by
\begin{equation}
\sigma_{\rm inelastic} = \sigma_{\rm elastic} \sqrt{1-\frac{2 \delta}{\mu v^2}},
\end{equation}
where $\delta$ is the mass splitting between the dark matter and its excited state, $\mu$ is the dark matter-nucleus reduced mass, and $v$ is dark matter particle's velocity. The mass splitting, $\delta$, plays an important role in both direct detection experiments, and in the capture of dark matter particles onto the Sun. In particular, dark matter particles with a splitting of $\delta \approx 100$ keV and a mass of 100 GeV must have a velocity of $v\gsim 575$ km/s if they are to inelastically scatter off of an iodine nucleus ($A=127$).  Including the motion of the Solar System through the dark matter halo, a non-negligible fraction of the dark matter particles are expected to exceed this velocity, especially in the summer months, leading to the prediction of a high event rate with strong annual modulation in an experiment such as DAMA. In an experiment such as CDMS, which uses Germanium ($A=73$) and Silicon ($A=28$) target nuclei, however, the velocity threshold is significantly higher ($v\gsim 665$ km/s), leading to considerably suppressed event rates. This relative suppression of the scattering rate with light target nuclei is the essence of the motivation for inelastic dark matter.

In order for inelastic dark matter to avoid exceeding the limits of CDMS and XENON10, while also generating the annual modulation reported by DAMA, the scatterings must excite the dark matter into a state $\delta\approx 80-150$ keV more massive that the ground state~\cite{cdms,xenoninelastic}. As a result of this mass splitting, inelastic dark matter scattering of off nuclei in the Sun is highly suppressed.  In particular, essentially no scattering occurs in the Sun between inelastic dark matter and nuclei lighter than carbon. Although iron nuclei constitute only 0.16\% of the Sun's mass, they dominate the capture rate of inelastic dark matter in the Sun.

This suppression is not found, however, in the case of more dense stars, such as white dwarfs or neutron stars. The velocity of a dark matter particle that falls into a white dwarf from rest will exceed 5000 km/s at the surface, and 8000 km/s near the core (in contrast to approximately 1300 km/s near the core of the Sun). In this situation, $2 \delta/\mu v^2 \ll 1$, and the effects of the dark matter's mass splitting are negligible. For example, an inelastic dark matter particle falling from rest into a white dwarf consisting mostly of carbon or oxygen will inelastically scatter with a cross section within a few percent of the value found in the elastic limit. 

\begin{figure*}[!]
\begin{center}
{\includegraphics[angle=0,width=0.49\linewidth]{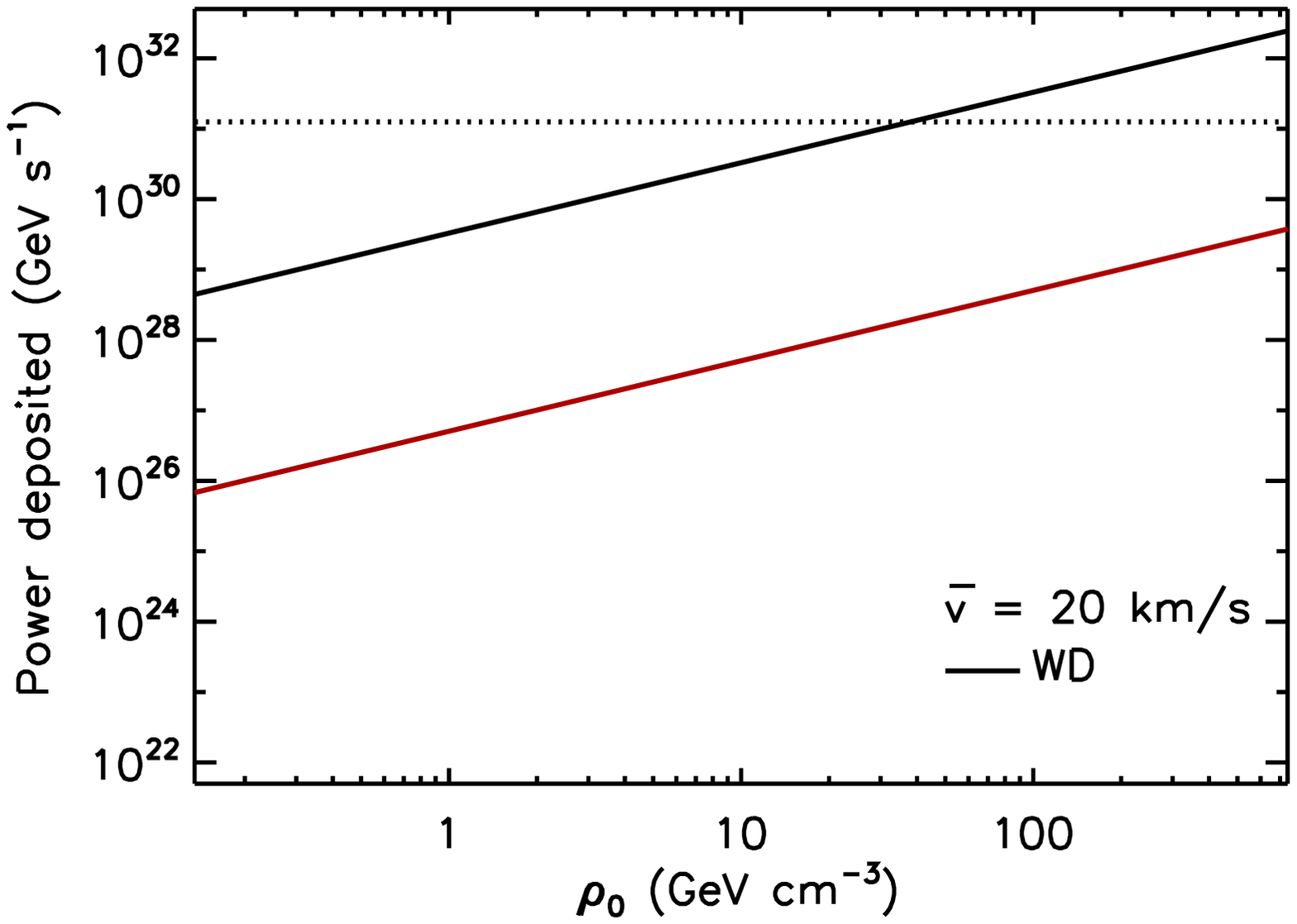}}
\hspace{0.1cm}
{\includegraphics[angle=0,width=0.49\linewidth]{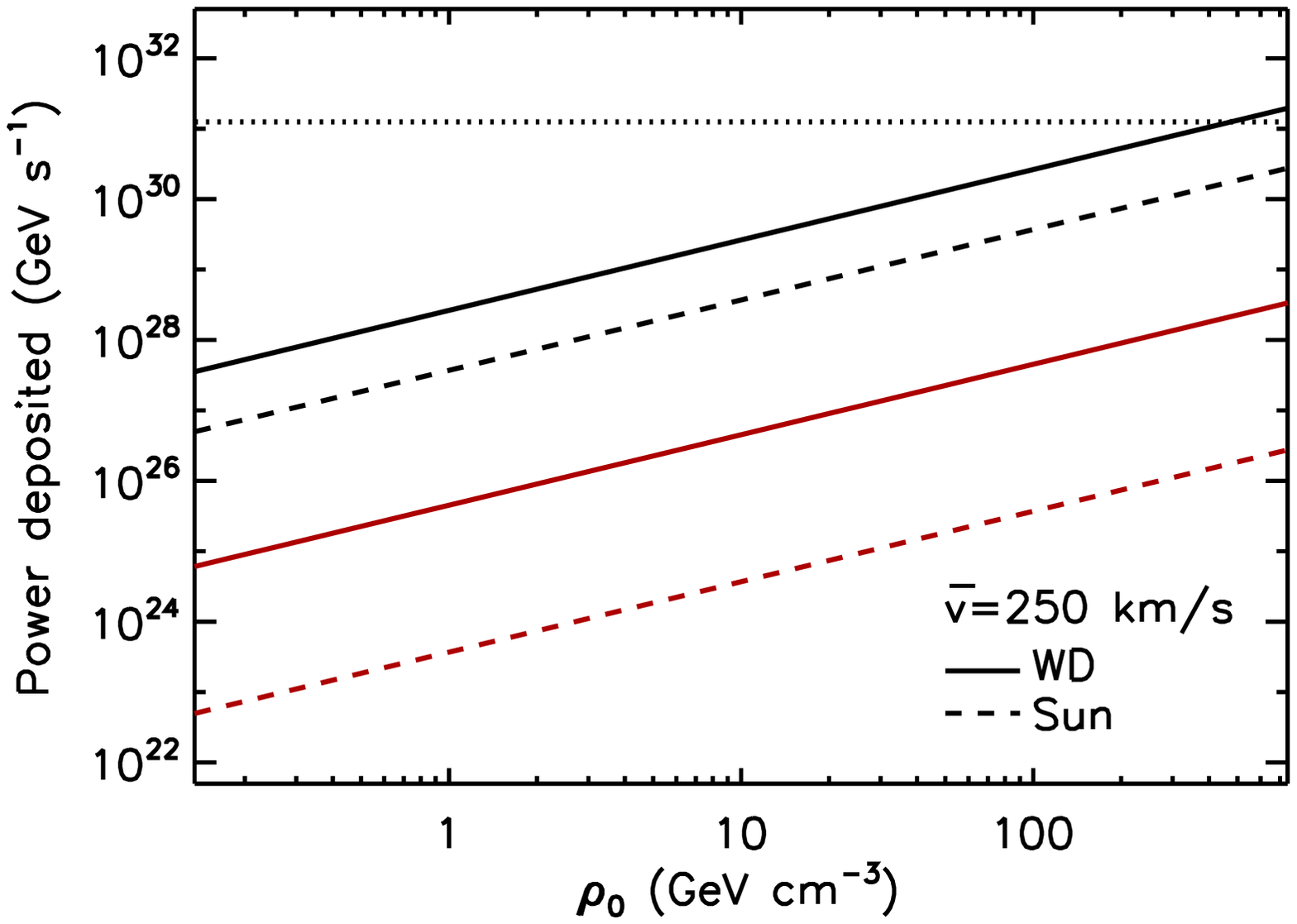}}\\
\vspace{-0.3cm}
\caption{The rate at which energy is injected into white dwarf stars (solid) and into the Sun (dashed) by annihilating inelastic dark matter, as a function of the environment's dark matter density. In the left frame, we show results for a dark matter velocity dispersion of 20 km/s (appropriate for globular clusters~\cite{clusteresc}), whereas in the right frame we use a velocity dispersion of 250 km/s and a star's velocity of 220 km/s (approprite for the Solar System). In each frame, upper (black) and lower (red) curves use dark matter-nucleon scattering cross sections of $10^{-40}$ cm$^2$ and $10^{-44}$ cm$^2$, respectively ($\sigma \approx 10^{-40}$ cm$^2$ is required to generate the DAMA annual modulation in the inelastic dark matter scenario, whereas $\sigma \approx 10^{-44}$ cm$^2$ is approximately the maximum value allowed for elastically scattering dark matter candidates). In each case, we have adopted a dark matter mass of 100 GeV, and a mass splitting of $\delta=100$ keV, although our results are not strongly sensitive to these values. The horizontal dotted lines denote the energy deposition rate above which observations of cool (3000-4000 K) white dwarfs would conflict with the predictions of inelastic dark matter.}
\label{fig}
\end{center}
\end{figure*}

To calculate the capture rate of inelastic dark matter particles in a white dwarf, we follow the approach described in Refs.~\cite{weiner,wang} for the case of capture in the Sun. For the Sun, we find results which are in good agreement with these previous studies (see the dashed lines in the right frame of Fig.~\ref{fig}). 

White dwarf stars can be well modeled as a system in hydrostatic equilibrium with an equation of state of a cold fermi gas.  Following Refs.~\cite{kippenhahn,webpage}, we describe the equation of state of a white dwarf as
\begin{eqnarray}
P&=&C_1 f(x),\nonumber \\
\rho&=&C_2 x^3, \nonumber \\   
x&=&k_f/(m_e c),
\end{eqnarray}
where $P$ is the pressure, $\rho$ is the density of the star, and $k_f$ is the Fermi momentum. The numerical coefficients and the function $f(x)$ are given by
\begin{eqnarray}
C_1&=&\frac{\pi m_e^4 c^5}{3h^3}, \nonumber \\   
C_2&=&\frac{ 8\pi m^3_e c^3}{3h^3},\nonumber \\
f(x)&=&x(2x^2-3)(x^2+1)^{1/2}+3 \ln [x+(1+x^2)^{1/2}], \,\,\,\,\,\,\,\,\, 
\end{eqnarray}
where $h$ is the Planck constant. Substituting into the above equations for pressure and density, the equation of hydrostatic equilibrium becomes

\begin{equation}
\frac{C_1}{C_2} \frac{1}{r^2} \frac{d}{dr} \bigg(\frac{r^2}{x^3}\frac{df(x)}{dr}\bigg)=-4\pi G C_2 x^3,
\end{equation}
which we solve numerically using a publicly available code~\cite{webpage} that also includes corrections due to electrostatic~\cite{Hamada} and general relativistic effects.  This approach gives an accurate description of the stellar structure of white dwarfs and fits the data exquisitely (see, for example, Fig.~3 of Ref.~\cite{webpage}).

One way in which the calculation of inelastic dark matter capture by white dwarfs departs from that described in Refs.~\cite{weiner,wang} is in the treatment of the star's optical depth. In particular, for cross sections as large as are required to produce the DAMA annual modulation ($\sigma$\,$\sim$\,$10^{-40}$ cm$^2$), dark matter particles are generally expected to scatter many (typically several tens of) times as they pass through a white dwarf. In this optically thick limit, the rate at which dark matter particles are captured simplifies considerably to~\cite{thick,bertone}
\begin{equation}
\Gamma_c \approx \bigg(\frac{8 \pi}{3}\bigg)^{1/2} \, \frac{3G \, R_{\rm WD}\,  M_{\rm WD}\, \rho_{\rm DM}}{m_{\rm DM}\, \bar{v}},
\end{equation}
where $R_{\rm WH}$ and $M_{\rm WD}$ are the radius and mass of the white dwarf, respectively, $\rho_{\rm DM}$ is the density of dark matter in the star's environment, and $\bar{v}$ is the dark matter's velocity dispersion (prior to infall into the gravitational potential of the star). Throughout our study, we will adopt $M_{\rm WD}=0.7 M_{\odot}$ and $R_{\rm WD}=0.0093 R_{\odot}$ as our default values (the mass distribution of white dwarfs is strongly peaked near 0.5-0.7 $M_{\odot}$~\cite{mass}). Note that as long as we remain within the optically thick limit, the capture rate does not depend on the interaction cross section of the dark matter, but only on its number density and velocity distribution.

The large cross section required in the inelastic dark matter scenario quickly leads to an equilibrium between the rates of capture and annihilation in a white dwarf, assuming that the dark matter particles are capable of self-annihilation (which might not be true if the dark matter carries a conserved quantum number rather being stabilized by a parity, for example). As a result, the capture rate can be converted into the rate at which additional energy is injected into the star's core. In a typical white dwarf star ($R_{\rm WD} \approx 0.0093 R_{\odot}$ and $M_{\rm WD}\approx 0.7 M_{\odot}$), inelastic dark matter annihilations provide a contribution to the star's luminosity that is given by:
\begin{eqnarray}
L &\approx& \Gamma_c \, m_{\rm DM} \nonumber \\
&\approx& 2.5\times 10^{28} \,{\rm GeV/s} \,\bigg(\frac{\rho_{\rm DM}}{1 \,{\rm GeV/cm}^3}\bigg) \bigg(\frac{220 \, {\rm km/s}}{\bar{v}}\bigg)\nonumber \\
&\approx& 4\times 10^{25} \,{\rm erg/s} \,\bigg(\frac{\rho_{\rm DM}}{1 \,{\rm GeV/cm}^3}\bigg) \bigg(\frac{220 \, {\rm km/s}}{\bar{v}}\bigg).
\label{power}
\end{eqnarray}

In Fig.~\ref{fig}, we show the rate at which inelastic dark matter deposits energy into a white dwarf (or into the Sun), as a function of the environment's dark matter density. We show results for dark matter with velocity dispersions of 20 km/s (left) and 250 km/s (right), and use two values for the dark matter-nucleon cross section: $10^{-44}$ cm$^2$ (lower red curves) and $10^{-40}$ cm$^2$ (upper black curves). In the case of the larger (smaller) cross section choice, a white dwarf is within the optically thick (thin) limit. As a cross section of $\sigma \approx 10^{-40}$ cm$^2$ is required to generate the DAMA modulation, we will focus on the optically thick case in the remainder of the study. 

The coldest and least luminous observed white dwarfs have temperatures as low as 3000 K and luminosities of $(2-3) \times 10^{28}$ erg/s (following from the Stefan-Boltzmann law, a 3000 K blackbody of radius 0.01 $R_{\odot}$ is predicted to have a luminosity of $2.8\times 10^{28}$ erg/s). From these observations, we can conclude that any additional source of energy (such as dark matter annihilations) must not exceed approximately $2\times 10^{28}$ erg/s. Within the context of inelastic dark matter, this translates to an approximate limit of $\rho_{\rm DM} \lsim 500$ GeV/cm$^3$ ($\rho_{\rm DM} \lsim 40$ GeV/cm$^3$) for $\bar{v}=250$ km/s ($\bar{v}=20$ km/s). Although these density limits are orders of magnitude higher than the average dark matter density in the local region of the Milky Way ($\rho_{\rm DM,local}\approx$ 0.3-0.5 GeV/cm$^3$~\cite{ullio}), there exist regions that are thought to contain sufficiently high quantities of dark matter (and/or with sufficiently slow velocity distributions) to make white dwarf stars useful probes of inelastic dark matter. We discuss some of these possibilities in the following section.

\section{Dark Matter Densities In Regions Containing White Dwarf Stars}

Many of the oldest (and coldest) known white dwarf stars are located in the dense collections of stars known as globular clusters. If high densities of dark matter were also present in such systems, these very old and cool white dwarfs would provide an ideal environment to search for the effects of inelastic dark matter. Despite the very high densities of baryonic matter in globular clusters (often containing hundreds of thousands of stars within a volume smaller than a cubic parsec), however, there is no compelling dynamical evidence that these systems contain significant quantities of dark matter.

Within the modern paradigm of globular cluster formation, these objects form in the disk of galaxies rather than in the center of dark matter halos (in contrast to dwarf spheroidal galaxies, for example). Indeed, observations of the Antennae Galaxy dramatically demonstrate the formation of globular clusters within its disk. Regardless, the formation of a globular cluster is generally expected to enhance the dark matter density through adiabatic contraction. In particular, the orbits of dark matter particles that are gravitationally bound to the cluster become contracted as the baryons ({\it ie.}~the stars) themselves gather more densely. This can lead to a contracted dark matter density in the globular cluster's core approximately given by
\begin{equation}
\rho_{\rm DM,core} \approx \rho_{\rm b,core}\, \bigg(\frac{\rho_{\rm DM,i}}{\rho_{b,i}}\bigg) \bigg(\frac{\rho_{\rm ISM}}{\rho_{\rm core}}\bigg)^{1/4}, 
\end{equation}
where $\rho_{\rm b,core}$ is the density of baryons in the globular cluster's core, $\rho_{\rm DM,i}$ is the initial density of {\it gravitationally bound} dark matter particles prior to adiabatic contraction, $\rho_{b,i}$ is the initial density of baryons prior to adiabatic contraction, $\rho_{\rm ISM}$ is the total density of the interstellar medium, and $\rho_{\rm core}$ is the total density in the core of the cluster. This result was found analytically using the standard adiabatic contraction model of Blumenthal {\it et al.}~\cite{blumenthal}, and was verified numerically using the adiabatic contruction code, \emph{Contra}~\cite{contra}.

Considering as an example the nearby globular cluster M4, we would expect that within an initial distribution of dark matter with a velocity dispersion of $\bar{v}\sim 220$ km/s, only $\sim 10^{-3}-10^{-4}$ or less of the particles will be gravitationally bound, thus significantly limiting the impact of adiabatic contraction. After adiabatic contraction, we estimate that the dark matter density in the core of M4 could be enhanced to on the order of a few GeV/cm$^3$ which, although an order of magnitude greater than prior to the formation of the cluster, is well below the density needed to significantly effect white dwarf stars.

As an alternative to globular clusters, we also consider white dwarfs within the dark matter halos of dwarf spheroidal galaxies. Dwarf spheroidals are highly dense, dark matter dominated systems, and contain very old stars~\cite{old}, making them excellent environments in which to study the impact of inelastic dark matter on white dwarfs. Ref.~\cite{list}, for example, describes seven dwarf spheroidals (Carina, Draco, Fornax, Leo I, Leo II, Sculptor, and Sextans) which have dark matter densities of 40-150 GeV/cm$^3$ and velocity dispersions of $\sim$10-20 km/s within their inner 10-20 parsecs. If old and cold white dwarf stars are present in such environments, we predict that inelastic dark matter will inject energy into them at a rate rivaling their luminosity. As a result, the observation of a white dwarf in such an environment with a temperature below approximately 4000 K would conflict with the inelastic dark matter hypothesis. Although dwarf spheroidals are typically very distant, and thus challenging to observe, there are 3-4 known dwarf spheroidals (Segue I, Segue II, Ursa Major II, and possibly Wilman I) within $\sim$30 kpc of the Solar System, making them within the approximate range of a 30 meter telescope~\cite{30m}. In the more distant future, a 100 meter telescope could place white dwarf stars in all Milky Way dwarf spheroidals within reach.

Lastly, the dark matter density is generally predicted to increase as one approaches the center of the Milky Way. According to the frequently used Navarro-Frenk-White (NFW) profile~\cite{nfw}, for example, the smooth component of the dark matter density in the Milky Way can be parameterized by
\begin{equation}
\rho_{\rm DM} \propto \frac{1}{r [1+(r/R_s)]^2},
\end{equation}
where $r$ is the distance to the Galactic Center, and $R_s\approx 20$ kpc is the scale radius. In addition, the velocity dispersion is predicted to decrease according to $\sigma_v \propto (r/R_s)^{1/2}$. Combining these effects, we estimate that the rate at which inelastic dark matter will accumulate in a white dwarf located at 1 kpc (100 pc) from the Galactic Center is $\sim$$40$ ($\sim$$1250$) times greater than the rate predicted in the vicinity of the Solar System. Adopting an NFW profile, we expect a sizable fraction of the luminosity of cold white dwarf stars within the inner $\sim$100 pc of the Milky Way to originate from inelastic dark matter annihilations. This effect can be considerably enhanced if the Milky Way's dark matter halo were adiabatically contracted by baryons~\cite{ac}. We therefore expect future infrared surveys of the inner Milky Way to provide a valuable test of inelastic dark matter.

\section{Conclusions}

In this paper, we have studied the capture and subsequent annihilation of inelastic dark matter particles in white dwarf stars. We find that inelastic dark matter (as motivated to generate the annual modulation claimed by DAMA/LIBRA without exceeding the limits from other direct detection experiments) is predicted to be captured by compact stars at a rate multiple orders of magnitude higher than elastically scattering dark matter. This makes inelastic dark matter a very efficient source of energy for white dwarf stars, potentially leading to observable consequences.

The role of inelastic dark matter is most pronounced in those stars that reside in regions with a high density (and/or low velocity dispersion) of dark matter.  In particular, we find that very old white dwarf stars in the inner tens of parsecs of dwarf spheroidal galaxies will be powered primarily by inelastic dark matter annihilations, preventing their otherwise expected cooling. This is also predicted to be the case for old white dwarfs found within the inner kiloparsec of the Milky Way. If future surveys were to discover very cold (T $\lsim$ 4000 K) white dwarfs in environments such as these, it would disfavor the inelastic dark matter hypothesis. In contrast, the lack of such stars discovered in future observations could be used to bolster the case for inelastic dark matter.

\bigskip

{\it Acknowledgements:} While we were in the final stages of this project, Ref.~\cite{fair} appeared on the LANL archive. While our numerical results are in good agreement, we have chosen to focus our study on inelastic dark matter captured by white dwarfs in dwarf spheroidal galaxies, and in the inner Milky Way, rather than in globular clusters which we have argued are unlikely to contain high densities of dark matter. The authors are supported by the US Department of Energy, including grant DE-FG02-95ER40896. DH is also supported by NASA grant NAG5-10842.%

\end{document}